\documentclass[a4paper]{jacow}

\usepackage{authblk}
\usepackage{graphicx}

\begin{document}

\title{Synchrotron radiation interaction with cryosorbed layers for astrochemical investigations}

	 \author[1]{R. Dupuy}
	 \author[1]{M. Bertin}
	 \author[1]{G. F\'{e}raud}
	 \author[1]{X. Michaut}
	 \author[1]{T. Putaud}
	 \author[1]{L. Philippe}
	 \author[1]{P. Jeseck}
	 \author[2]{R. Cimino}
	 \author[3]{V. Baglin}
	 \author[4]{C. Romanzin}
	 \author[1]{J.-H. Fillion} 
	 
	 \affil[1]{Sorbonne Universit\'e, Observatoire de Paris, Universit\'e PSL, CNRS, LERMA, F-75005, Paris, France}
	 \affil[2]{Laboratori Nazionali di Frascati (LNF)-INFN I-00044 Frascati}
	 \affil[3]{CERN, CH-1211 Geneva 23, Switzerland}
	 \affil[4]{Laboratoire de Chimie Physique, CNRS, Univ. Paris-Sud, Universit\'e Paris-Saclay, 91405, Orsay, France}

	\date{}

\maketitle

\abstract{Photon-stimulated desorption (PSD) is a process of interest for the two seemingly unrelated topics of accelerator vacuum dynamics and astrochemistry. Here we present an approach to studying PSD of interstellar ice analogs, i.e. condensed films of molecules of astrophysical interest at cryogenic temperatures, using synchrotron radiation. We present results obtained in the VUV range on various pure and layered ices, focusing on elucidating the desorption mechanisms, and results in the X-ray range for H$_2$O.}

\section{Introduction}

Photon-stimulated desorption, the process by which an adsorbed molecule on a surface is detached by a photon, is one example of a surface science process whose study has applications in different fields. In the context of accelerators, the synchrotron radiation generated by the circulating beams of charged particles hits the walls and releases molecules into the gas phase, thus limiting the vacuum and potentially hindering beam stability. During beam operations (in the presence of photons, electrons and energetic ions), the vacuum in an accelerator becomes orders of magnitude higher owing to non-thermal processes, which are thus completely dominant\cite{turner1999,brandt2007}. Studies of photon-stimulated desorption and its impact on the vacuum dynamics of accelerators have been made in this context. 

Studies of photodesorption have also been made in a very different context: astrochemistry of the cold interstellar medium (see \cite{westley1995,oberg2009a,yabushita2013,arasa2015,martin-domenech2015,cruz-diaz2017,dupuy2017a} and references therein). The cold interstellar medium and accelerators present interesting similarities in this context. As will be detailed in part 2, the cold interstellar medium also features very low densities, molecules (similar to those composing the residual gases of vacuum chambers) adsorbed on low temperature surfaces (relevant for accelerators such as the LHC operating at cryogenic temperatures), and the presence of photons and energetic particles. 

The general context of astrochemistry will be briefly introduced, then our approach to the study of photodesorption, which makes use of synchrotron radiation to uncover fundamental mechanisms, will be presented in both the VUV photon range and in the more recently explored X-ray photon range. 

\section{Astrochemical context}

In the interstellar medium, molecules are found in many different regions \cite{tielens2013}. Observational spectroscopy spanning from the centimeter to the far UV region of the spectrum allowed the identification of more than 200 molecules. Radio-astronomy, which identifies molecules through their rotational lines, has probably been the most fruitful method for identifying molecules so far. Aside from gas phase molecules, in all these regions are also found tiny, carbonaceous and/or silicated dust grains ($<$1 $\mu$m), which can act as reaction catalysts for the formation of molecules. Of particular interest here are the cold ($<$100K) and dense ($>$10$^3$ molecules.cm$^{-3}$) regions. These regions typically start off as so-called molecular clouds. Some parts of these molecular clouds can form denser clumps which through gravitational collapse end up more and more dense, and also colder and colder as they become completely shielded from external irradiation. These "dense cores" then evolve as the gravitational collapse forms a protostar, with its protostellar envelope. Next is the formation of a young star, with creation of a surrounding disk of matter called a protoplanetary disk, which will eventually evolve into a system of planets, asteroids and comets \cite{vandishoeck2017}. 

In all these stages, the aforementioned dust grains can be cold enough to act as molecular sinks: molecules that form at their surface or that accrete from the gas phase stay cryosorbed. Thus in parallel of the gas phase, there is a solid phase, usually called the ice mantles. The typical content of these ice mantles is known mainly thanks to mid-infrared spectroscopy, although it is much harder to constrain than the content of the gas phase. The main component by far is H$_2$O. Following are CO and CO$_2$ with a few tens of percent with respect to H$_2$O, and CH$_3$OH, NH$_3$, CH$_4$ and a few molecules at the few percent level. These numbers can vary a lot depending on the observed source \cite{boogert2015}. 

In regions where these grains are exposed to heat, the interaction between this solid phase and the gas phase can happen through thermal desorption of molecules. However, when the temperature stays cold, this interaction can only happen through non-thermal desorption mechanisms. In these cold regions, observations of molecules in the gas phase that could only have formed on the grains, or of molecules that should be completely frozen out on the grains and absent from the gas, showed that such mechanisms should be at play. Energy coming from cosmic rays, exothermic chemistry, shocks or photons can cause desorption of molecules. This is why photon-stimulated desorption is of interest here.  

The importance for astrochemistry of the photodesorption process and of its study with dedicated laboratory experiments has been demonstrated by astrochemical models implementing it. Astrochemical models aim to help the interpretation of astronomical observations of molecules and extract meaningful information from them. They include a physical modeling of the considered region and a chemical network of up to thousands of reactions to try to reproduce the observed abundances of molecules. For example, in a typical model of protoplanetary disk \cite{kamp2013}, if the photodesorption process is turned off, the prediction is that there is no water in the gas phase in the cold, outer part of the disk, which contradicts observations \cite{hogerheijde2011}. The inclusion of photodesorption is necessary to reproduce the observations. Another example in protoplanetary disks is the detection of CH$_3$CN, the first moderately complex organic molecule to be detected in this kind of astrophysical object \cite{oberg2015a}. The current hypothesis is that this molecule forms at the surface of cold grains, and is then released by photodesorption in the gas phase where it is detected. 

In order to implement the photodesorption process in astrochemical models, laboratory studies are mandatory to obtain quantitative inputs (number of desorbed molecules per incident photon) and their dependence on various parameters (most importantly the nature of the molecule, but also the temperature, ice morphology, thickness, etc...). 

\begin{figure}
    \includegraphics[trim={1cm 7.2cm 16cm 1cm},clip,width=\linewidth]{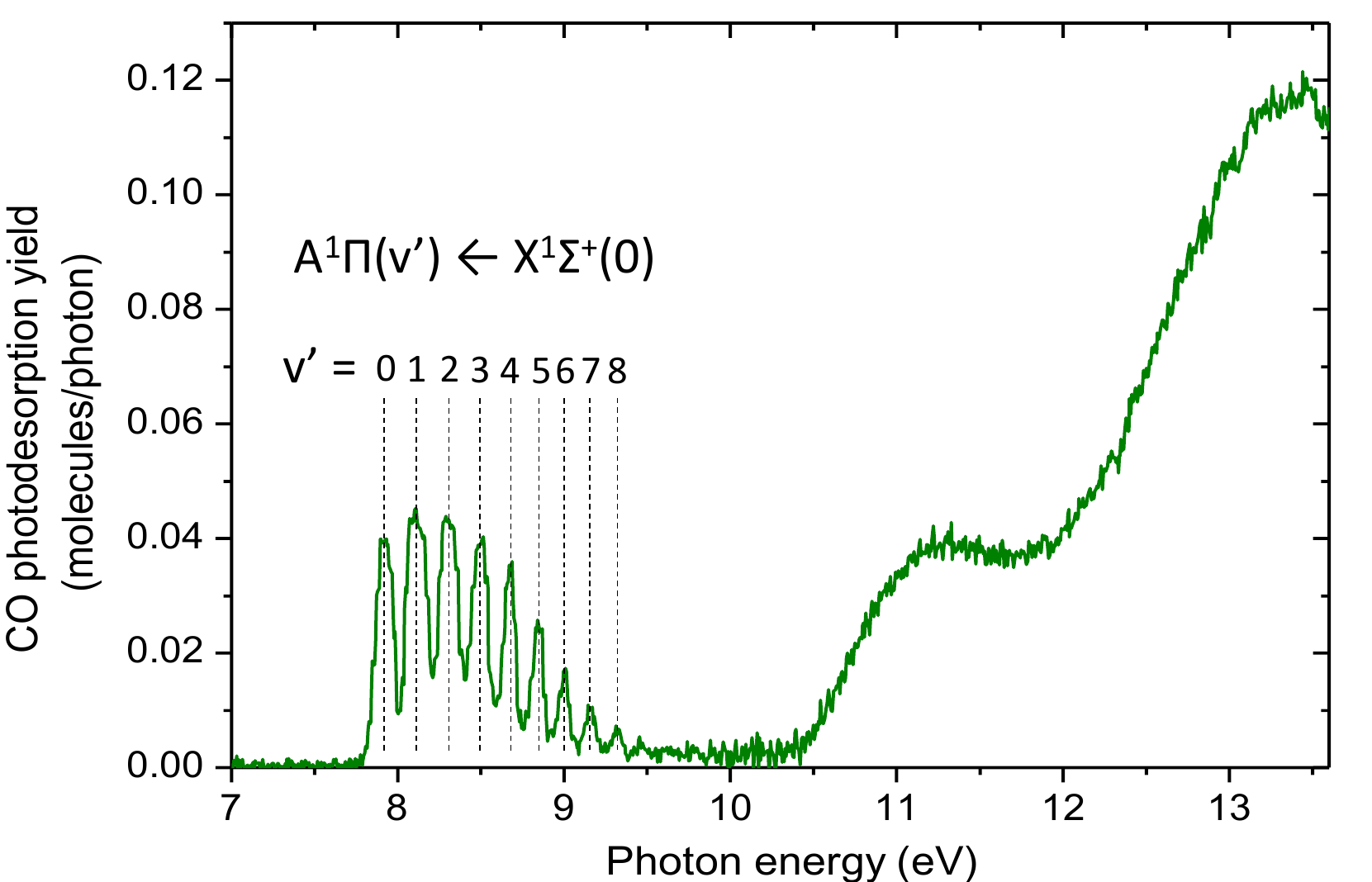}
    \caption{Photodesorption yield of CO as a function of photon energy, for a 20 ML thick CO ice at 10 K. Adapted from ref \cite{dupuy2017}}
    \label{CO}
\end{figure}   

\section{Methods}

The experiments were performed in the SPICES set-up and its upgraded version, the SPICES 2 set-up. These set-ups are described in more detail elsewhere \cite{doronin2015, dupuy2018}. Briefly, they consist of an ultra-high vacuum chamber with a rotatable sample holder cooled down to 10-20 K by a closed-cycle helium cryostat. The sample holder has several surfaces that can be used. The experiments were typically performed on either polycristalline gold or OFHC copper. In SPICES 2, a copper surface is electrically isolated from the sample holder, allowing to measure the current at the sample, which is used for total electron yield measurements during X-ray irradation. A gas injection line connected to a dosing tube inside the chamber allows to deposit molecular films of controlled thicknesses \cite{doronin2015} (0.2 - 200 monolayers). The films deposited can be probed using an infrared spectrometer (RAIRS-FTIR). During irradiation, the desorption of molecules is monitored by a quadrupole mass spectrometer. In SPICES 2, two different mass spectrometers can be used, one dedicated to the detection of neutral species, and the second one also being able to detect positive and negative ions and to filter them as a function of their kinetic energy using a 45$^{\circ}$ deflector. 

The set-ups are mobile and can be connected, among other possibilities (lasers, etc) to beamlines at the SOLEIL synchrotron. Experiments in the VUV range were performed at the DESIRS beamline, and experiments in the X-ray range at the SEXTANTS beamline. On the DESIRS beamline the output of the undulator is monochromatized with a resolution of $\sim$0.04 eV and scanned continuously between 7 and 14 eV. The second harmonics is suppressed by a Krypton gas filter. The typical flux in this mode is 10$^{13}$ photons.s$^{-1}$. On the SEXTANTS beamline, similarly the output is monochromatized with a resolution of $\sim$0.15 eV and scanned continuously between 520 and 600 eV, around the O 1s edge. The flux is about 10$^{13}$ photons.s$^{-1}$. More details can be found elsewhere on the beamlines \cite{nahon2012,sacchi2013} and their use in our experiments \cite{fayolle2011,dupuy2018}. 

The results we obtain are "photodesorption spectra", which is the desorption signal of a given molecule as a function of photon energy. The signal on the relevant mass channel of the QMS, which is proportional to the desorption flux of the molecule, is divided by the photon flux. This is then calibrated to an absolute number of molecules desorbed per incident photon \cite{fayolle2011,dupuy2018}. More details on the specific experiments mentioned later can be found in the corresponding papers (see citations below). 

\section{VUV photodesorption}

\begin{figure}
    \includegraphics[trim={4cm 0cm 17.5cm 2.5cm},clip,width=\linewidth]{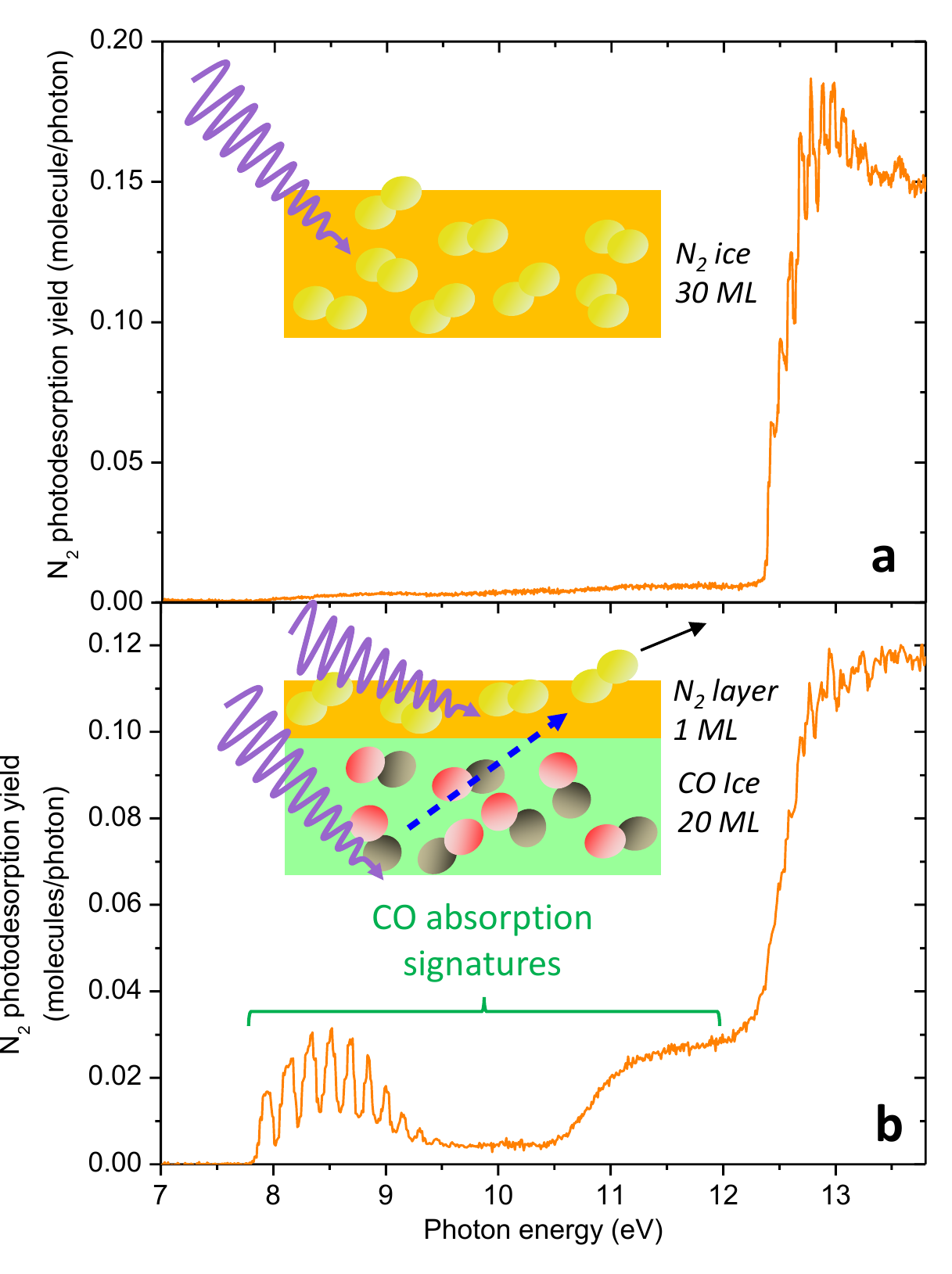}
    \caption{a. Photodesorption yield of N$_2$ as a function of photon energy, for a 30 ML thick N$_2$ ice at 18 K. b. Photodesorption yield of N$_2$ as a function of photon energy, for 1 ML of N$_2$ layered on a 20 ML CO ice. Adapted from ref \cite{fayolle2013}}
    \label{N2}
\end{figure}

A prototypical example of the kind of information obtained through the photodesorption spectrum of a molecule, thanks to synchrotron radiation, is the CO molecule. Its photodesorption spectrum is shown in fig. \ref{CO}. The spectrum is taken between 7 and 13.6 eV. The upper limit of 13.6 eV comes from the astrophysical context: in astrophysical media photons above 13.6 eV are absorbed by atomic hydrogen, which is by far the most abundant element, and thus they do not need to be considered. The photodesorption yield varies wildly as a function of photon energy, and the spectrum is very structured. These structures can be readily identified with the electronic transitions of condensed CO: for example, the peaks between 8 and 9 eV correspond to the vibrational progression of the A-X transition. These molecular signatures give a first basic information, which is that the initial step of desorption is the absorption of the photon by a CO molecule (and not e.g. the substrate), and that a single-photon process is involved (otherwise, for multi-photon processes, the structures would be deformed when compared to the absorption spectrum). This is usually termed Desorption Induced by Electronic Transitions (DIET). How the electronic energy is converted into a desorption event is the more difficult question. More details can be found on CO photodesorption in ref. \cite{fayolle2011}.

The signal on the mass spectrometer allows to identify which species desorbs in the end, while the spectral information tells which species initially absorbed a photon. The desorbed molecule need not be the same as the one which absorbed a photon, however. This can be shown for a system with different species, such as a layer of one molecule on a substrate of another molecule. It has been shown by layering isotopically labeled $^{13}$CO on a $^{12}$CO ice that the desorbed molecules are the surface-most $^{13}$CO molecules, but that they can be desorbed following absorption of photons by sub-surface $^{12}$CO molecules, as the spectral signatures of the isotopes slightly differ \cite{bertin2012}. The result is even more striking for N$_2$ layered on a CO ice \cite{bertin2013}. Fig \ref{N2}a shows the photodesorption spectrum of pure N$_2$ ice. The first allowed electronic transition of N$_2$ is rather high in energy and thus strong photodesorption is only observed above 12 eV. When one monolayer of N$_2$ is layered on top of a CO ice, however, monitoring the desorption of N$_2$ we can see appearing in the spectrum peaks between 8 and 9 eV that are the molecular signature of CO absorption (fig \ref{N2}b). While a N$_2$ molecule desorbs, the photon was initially absorbed by a CO molecule. Thus an energy transfer of some sort occurs. This process has been termed "indirect DIET" before \cite{bertin2012}. 

\begin{figure}
    \includegraphics[trim={0.5cm 4.5cm 7.5cm 2cm},clip,width=\linewidth]{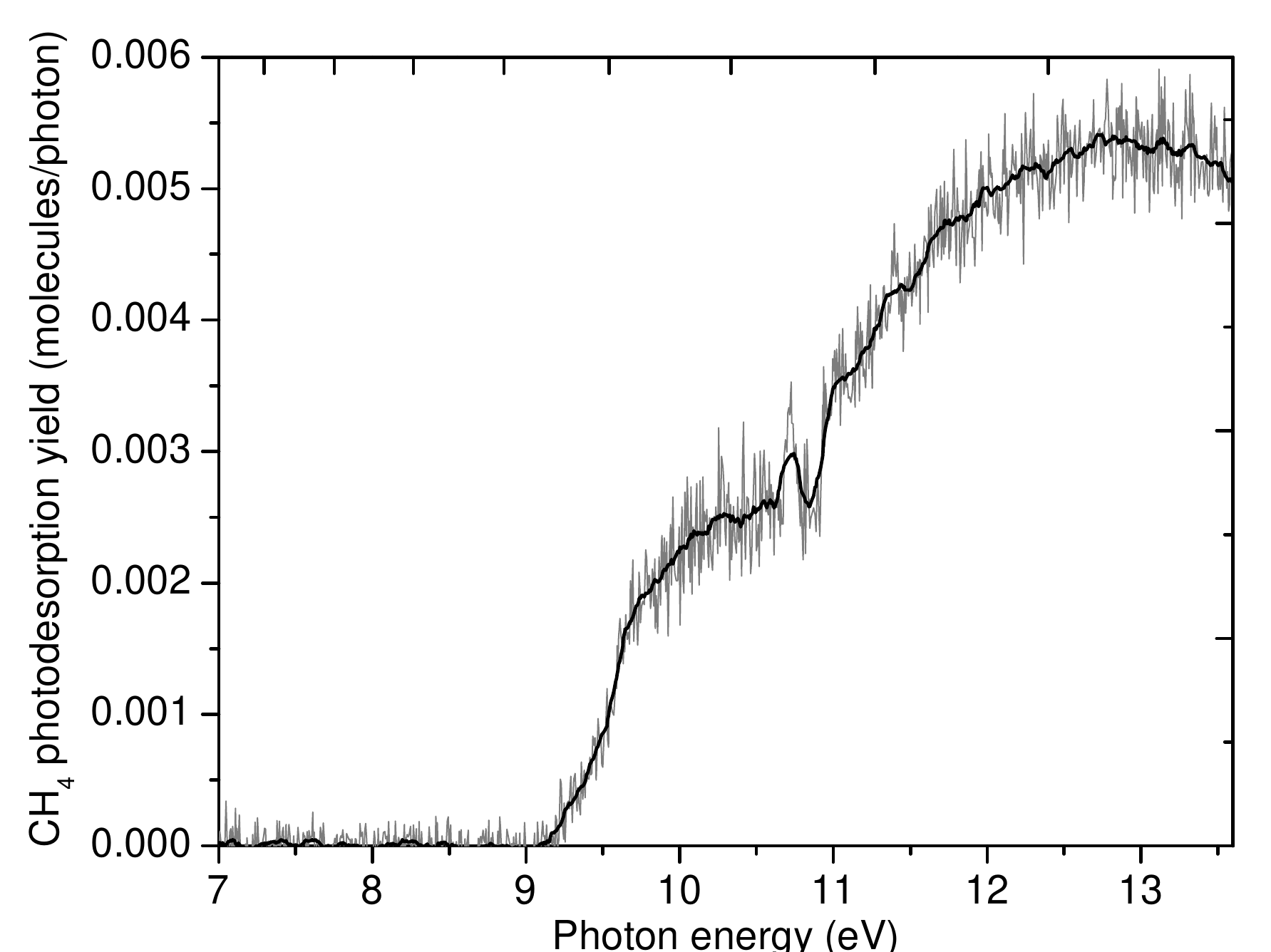}
    \caption{Photodesorption yield of CH$_4$ as a function of photon energy, for a 20 ML pure CH$_4$ ice at 10 K. Adapted from ref \cite{dupuy2017}}
    \label{CH4}
\end{figure}

The aforementioned molecules, CO and N$_2$, are not or not efficiently dissociated by VUV irradiation in the considered energy range. When molecules that dissociate are considered, many new possibilities for photodesorption open. Photochemistry and photodissociation can lead to the desorption of not just the parent molecule, but also other molecules and fragments, as in the case of methanol for example \cite{bertin2016}. For the parent molecule, new channels of desorption are also opened. The example taken here is the CH$_4$ molecule (more details are found in ref \cite{dupuy2017}). The photodesorption spectrum of CH$_4$ is shown in fig. \ref{CH4} and features the electronic signature of the molecule, with the electronic states being this time dissociative. Dissociation of methane creates many fragments in the ice (CH$_3$, CH$_2$, H, etc) which can subsequently react. Reactions that lead back to the CH$_4$ molecules are usually exothermic, and the excess energy can be used for desorption (for example, the CH$_3$ + H $\rightarrow$ CH$_4$ liberates $\sim$4.5 eV). Another possibility is that the fragments themselves can be energetic: in particular, the light H fragments can carry a lot of kinetic energy and could "kick" CH$_4$ molecules out of the ice. The kick out mechanism was initially proposed and studied in simulations of the photodesorption of water \cite{andersson2008}.

These examples illustrate some of the complexity of photon-stimulated desorption processes: even in the case of a simple, pure ice like CO, the mechanism of desorption is not exactly well known, and as soon as different molecules and chemistry are involved the degree of complexity increases. Indirect desorption changes the picture of photodesorption in a realistic astrophysical ice, which contains many molecules.

\section{X-ray photodesorption}

The UV range is particularly interesting for astrophysics as UV photons up to 13.6 eV are ubiquitous in the interstellar medium. But some regions, such as protoplanetary disks, also see considerable X-ray irradiation. Photodesorption by X-rays in an astrophysical context has been little studied so far and it is not yet taken into account in models. Therefore we performed experiments on X-ray photodesorption from pure water ice, for which some results are presented here. More details can be found on both the results and the astrophysical context in ref \cite{dupuy2018}. 

X-ray photodesorption, like UV photodesorption, is initiated by an electronic transition, except instead of exciting valence electrons, the core electrons of a molecule are excited. In the case of water the O 1s electrons are excited, in the 520-600 eV range. A core excitation for low atomic number atoms relaxes almost entirely through Auger decay. Auger decay leaves the molecule in a highly excited state that can give rise to dissociation pathways not accessible for single electron valence excitations. The desorption of some species will occur predominantly through such dissociation pathways. For other species, the dominant process will be the X-ray induced Electron Stimulated Desorption (XESD). The Auger electron, with about 500 eV of kinetic energy in the case of water, will be scattered in the ice, causing many secondary electrons and excitations. These secondary events, which are similar to the low-energy valence excitations in the VUV range, will be responsible for the desorption of certain species. In fact, since most of the energy of the initial photon goes in the XESD channel and creates many secondary events, XESD is expected to dominate the desorption of species which are also seen for low-energy excitations. 

\begin{figure}
    \includegraphics[trim={2cm 1cm 1cm 1cm},clip,width=\linewidth]{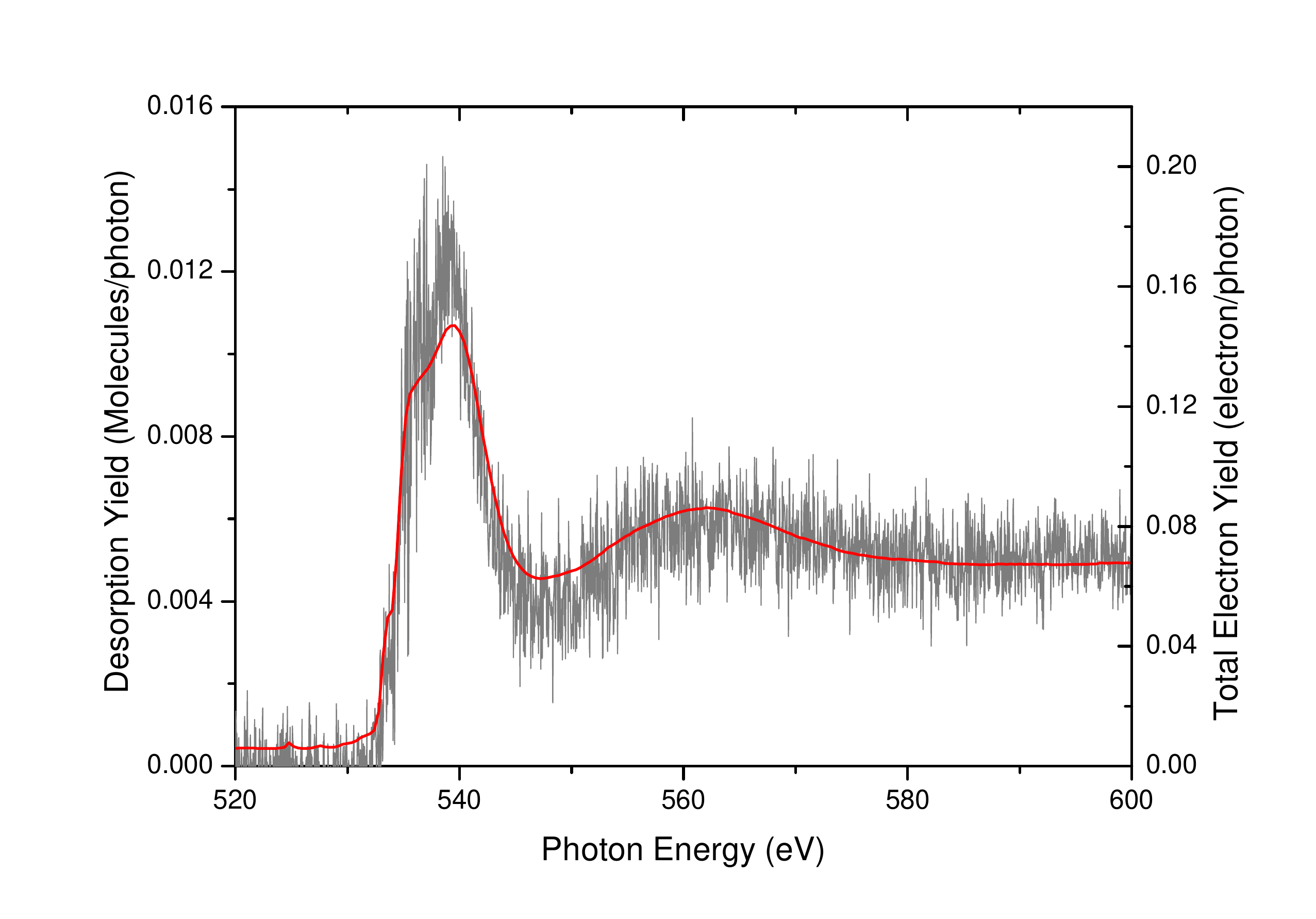}
    \caption{Photodesorption yield of H$_2$O for a 100 ML compact amorphous solid water (c-ASW) ice at 90K (grey trace). Also shown is the TEY (red trace). Adapted from ref \cite{dupuy2018}}
    \label{H2O_X}
\end{figure}

Fig. \ref{H2O_X} shows the photodesorption spectrum of the intact H$_2$O molecule around the O 1s edge, as well as the Total Electron Yield (TEY), i.e. the number of electrons escaping the surface per incident photon. The number of escaping electrons is proportional to the number of Auger decays occuring in the ice and thus to the number of absorbed X-ray photons. The TEY therefore represents the X-ray absorption spectrum (XAS). We can see that the photodesorption of H$_2$O follows the absorption. This is also the case for the other detected desorbing neutral species, O$_2$ and H$_2$ \cite{dupuy2018}. The contribution of the XESD mechanism, i.e. the electrons, to desorption should follow the total electron yield. The fact that the desorption spectra of neutrals follow exactly the TEY hints that XESD probably dominates. This is further compounded by the argument developed above, as neutral species are desorbed by low-energy excitation. The estimation of the photodesorption yield per absorbed photon of H$_2$O is also similar to the electron-stimulated desorption yield (see \cite{dupuy2018}), which is another argument for the dominance of XESD. 

\begin{figure}
    \includegraphics[trim={3cm 1cm 1cm 1cm},clip,width=\linewidth]{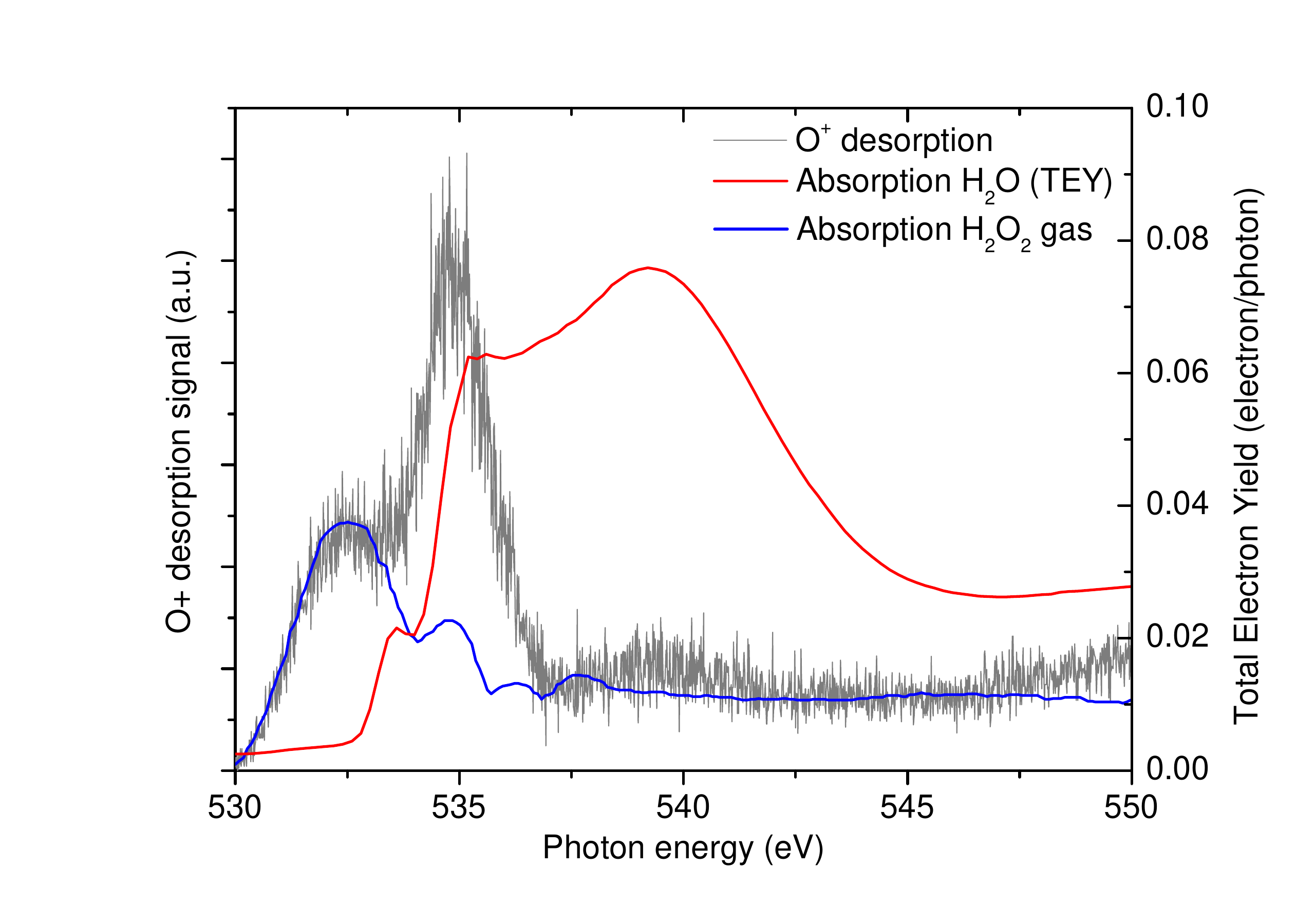}
    \caption{Photodesorption yield of O$^+$ for a 100 ML c-ASW ice at 90K (grey trace). Also shown are the TEY (red trace) and the gas phase absorption of H$_2$O$_2$ \cite{berkowitz2002} (blue trace), shifted to match the O$^+$ yield.}
    \label{O+}
\end{figure}

One example of a species whose desorption does not follow the TEY, and where XESD thus does not play a dominant role, is O$^+$. The photodesorption yield of O$^+$ is shown in fig. \ref{O+}. In fact, the features observed between 530 and 536 eV in the spectrum do not correspond to resonances of water: instead, the peaks can be identified to resonances of H$_2$O$_2$. This molecule is a product of the photolysis of water by the X-rays, present at the few percent level in the ice. The observations of the H$_2$O$_2$ peaks dominating the photodesorption spectrum of O$^+$ indicate that (i) O$^+$ is not dominated by secondary electrons and (ii) it is efficiently produced by direct excitation of H$_2$O$_2$, and not by excitation of H$_2$O. The differences in efficiency must be very large, as H$_2$O$_2$ is only a minor product in the irradiated ice, which is still mostly made of water. A similar behaviour is also observed for other O-bearing ionic fragments. One simple reasoning that can help explain this observation is the following: when dissociating a water molecule to form an oxygen-bearing fragment, an O-H bond (or two) are broken. In this dissociation process, the much lighter H or H$^+$ fragment will carry most of the kinetic energy, as the energy partition is expected to depend on the mass ratio of the fragments. Thus not enough kinetic energy is left for the oxygen-bearing fragment to desorb. Conversely, when dissociating H$_2$O$_2$, an O-O bond is broken, yielding a roughly similar amount of energy to both fragments, which then have the required energy to desorb. Confirming such an explanation will require further experiments. 

Besides these two species, many more are seen desorbing \cite{dupuy2018}. The neutral species are by far the most abundant desorption products, but as shown in the example of O$^+$, ion desorption can provide interesting information as well. Even in the case of a pure ice of a molecule like water, where the chemical network is not very complex, the number of different species desorbing is high and distinct desorption pathways exist. 

\section{Conclusion}

Photon-stimulated desorption by UV and X-ray photons is relevant for both astrochemistry and accelerator vacuum dynamics. Focusing here on multilayer physisorbed molecules, we showed that photodesorption yields depend strongly on (i) wavelength ($5\times10^{-2}$ to $< 1\times10^{-3}$ molecule/photon for CO over the VUV range) and (ii) molecules (a few $10^{-2}$ for CO to $< 1\times10^{-5}$ molecule/photon for methanol\cite{bertin2016}), but also depends on other factors like the presence of different molecules through indirect desorption. Comparing at the maxima of absorption in the VUV and X-ray ranges, X-ray desorption yields are higher (for H$_2$O, $1\times10^{-2}$ for X-rays to $3\times10^{-4}$ molecule/photon for UV). The variety of mechanisms that can be involved has been introduced.  

As shown with the example of photon-stimulated desorption, surface science studies can be of interest to vastly different fields like astrochemistry and accelerator vacuum dynamics. While the exact systems that are of interest for each field may differ, efforts to develop experimental techniques and knowledge can be shared. In particular, here we focused on attempts to unravel some fundamental mechanisms that can be more general than the particular examples presented, and from which conclusions can be drawn that are of broader interest. 

\bibliographystyle{IEEEtran}
\bibliography{Ecloud_proceedings}

\end{document}